\documentclass[aip,preprint]{revtex4-1}
\usepackage{graphicx}
\begin{document}


\title{High-resolution alternating-field technique to determine the magnetocaloric
effect of metals down to very low temperatures}


\author{Y. Tokiwa}
\email[]{ytokiwa@gwdg.de}
\author{P. Gegenwart}
\affiliation{I. Physik. Institut, Georg-August-Universit\"{a}t G\"{o}ttingen, D-37077 G\"{o}ttingen, Germany}


\date{\today}

\begin{abstract}
The magnetocaloric effect or "magnetic Gr\"uneisen ratio"
$\Gamma_H=T^{-1}(dT/dH)_S$ quantifies the cooling or heating of a
material when an applied magnetic field is changed under adiabatic
conditions. Recently this property has attracted considerable
interest in the field of quantum criticality. Here we report the
development of a low-frequency alternating-field technique for
measurements of the magnetocaloric effect down to very low
temperatures, which is an important property for the study of
quantum critical points. We focus in particular on highly conducting
metallic samples and discuss the influence of eddy current heating.
By comparison with magnetization and specific heat measurements, we
demonstrate that our fast and accurate technique gives
quantitatively correct values for the magnetocaloric effect under
truly adiabatic conditions.
\end{abstract}


\maketitle 

\section{Introduction}
Understanding the phenomena close to quantum critical points (QCPs)
is one of the major challenges in condensed matter physics. A QCP
emerges when the characteristic temperature of a system, e.g. the
N\'{e}el temperature, is suppressed to zero by variation of a
non-thermal parameter $r$ like pressure, magnetic field or doping.
Upon pressure-tuning a system towards a QCP, the Gr\"{u}neisen ratio
$\Gamma$ is expected to display singular behavior, namely a
divergence towards zero-temperature, while by tuning a system by
magnetic field towards quantum criticality $\Gamma_{\rm H}$, as
defined below, diverges.\cite{ZhuLijun:UnidGp,GarstM:SigctG} Both
Gr\"{u}neisen parameters consist of ratios of the control-parameter
and temperature derivatives of the entropy, which is accumulated
close to the QCP ($S$, $\alpha$, $C$ and $M$ denote the entropy,
volume thermal expansion, specific heat and magnetization,
respectively):
\begin{eqnarray}
&&\Gamma=-\frac{1}{V_mT}\frac{(\partial S/\partial p)_T}{(\partial S/\partial T)_p}= \frac{\alpha}{C_p}\\
&&\Gamma_{\rm H}=-\frac{1}{T}\frac{(\partial S/\partial
H)_T}{(\partial S/\partial T)_H}=-\frac{(\partial M/\partial
T)}{C_H}=\frac{1}{T}\left.\frac{dT}{dH}\right|_S \nonumber
\end{eqnarray}

Indeed, both $\Gamma$ and $\Gamma_H$ have been experimentally shown
to diverge in materials near
QCPs.\cite{KuchlerR:DivtGr,KuchlerR:Grurdq,Lorenz07,ISI:000263389500045}
Since the magnetic field, in contrast to pressure or doping, can
easily be tuned continuously in-situ, the magnetic Gr\"uneisen ratio
is ideally suited to investigate the nature of QCPs. So far, the
magnetic Gr\"uneisen ratio close to QCPs has been determined only
indirectly by calculation from magnetization and specific heat
data\cite{ISI:000263389500045}. Note, that $\Gamma_H$ can be
determined by a single measurement of the magnetocaloric (MCE)
effect, $(dT/dH)_S$, as shown above. Very recently, the
field-dependence of the entropy close to a QCP has been calculated
from highly non-adiabatic (quasi isothermal) $T(H)$ traces,
separating the effect of heat flow from/to the bath and the
magnetocaloric effect of the sample, using independent measurements
of the thermal conductance between sample and
bath.\cite{ARostScience09} However, a more direct measurement of the
MCE under truly adiabatic conditions is highly desired for
comparison with theoretical scenarios. Below, we show that the
alternating-field technique is suitable for this purpose.
Previously, the spin-flip transition in insulating GdVO$_4$ has been
investigated using such technique down to 0.5~K.\cite{Fischer} In
this study not much care has been taken on the absolute values of
the MCE and eddy current heating plays no role, since the material
is an electrical insulator. Below, we report the application of the
low-frequency alternating-field technique to clean metals and down
to mK temperatures. We carefully investigate the effect of eddy
current heating. Furthermore, we prove by comparison with
magnetization and specific heat measurements, that accurate absolute
values of the adiabatic MCE under are obtained.


\section{Experimental Setup}
The definition of the MCE implies that a change of the sample
temperature $\Delta T$ caused by a small change of the magnetic
field $\Delta H$ under adiabatic conditions needs to be detected.
Below, we demonstrate how the MCE could be detected directly by
applying a low-frequency alternating field, $H_0\sin(2\pi ft)$, and
detecting the temperature oscillation of the sample with the same
frequency, $T_0\sin[2\pi f(t+\delta t)]$ ($\delta t$ is a small
delay due to the thermal resistance between the sample and
thermometer).

Our setup for generating the alternating field is shown in Fig.~1. A
SR830 Lock-In amplifier (Stanford Research Systems) is used to
produce a current with low frequency (typically 0.01$\sim$0.1\,Hz,
the choice of the frequency will be discussed later). The current is
amplified by a Kepco bipolar power amplifier up to 8\,A and sent
through a modulation coil (8\,A current corresponds to 25\,mT) which
is inserted to the main superconducting magnet and cooled within the
liquid 4-He. The current is determined by measuring the voltage
across a shunt resistance (4\,m$\Omega$) with a Keithley 195A
multimeter. The modulation field is superimposed to the desired
static magnetic field of interest which is generated by a large
superconducting magnet. Note, that a finite static magnetic field is
required since the MCE depends on the magnetization of the system
(cf. eq. 1).

 \begin{figure}[t]
 \includegraphics{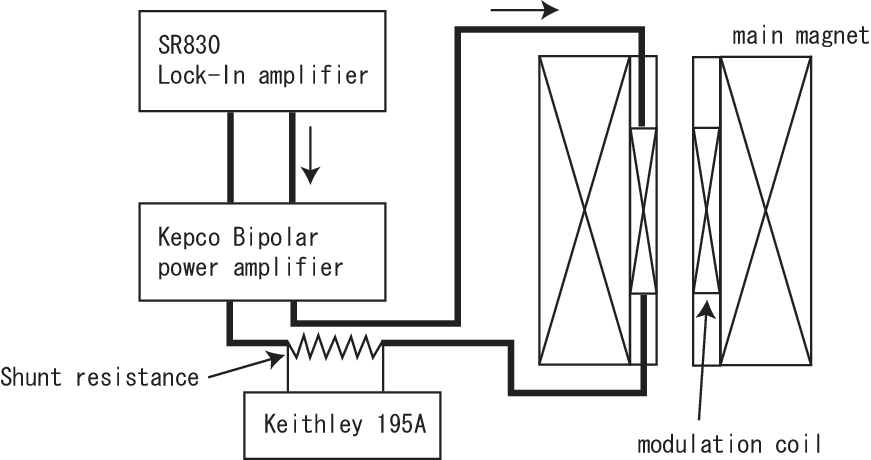}
 \caption{\label{}Generation and control of the alternating magnetic
  field superposed to the large magnetic field of the
  main superconducting magnet.}
 \end{figure}

Figure 2 shows the setup to detect the sample temperature under
quasi-adiabatic conditions. The sample is glued on a thin sapphire
plate and is cooled or heated through the weak thermal link (thin
CuNi wire). If $\tau$ denotes the thermal relaxation through the
link, the frequency of the field modulation must be adjusted such
that $1/f\ll\tau$ in order to obtain quasi-adiabatic conditions.
Such adjustment is easily be performed at several different
temperatures by taking temperature traces with differing frequency
and calculation of the MCE (see below). If the MCE is frequency
independent, the measurements are performed under quasi-adiabatic
conditions. The RuO$_2$ resistive thermometer is glued on the sample
and its leads are made from superconducting filaments to avoid an
additional heat leak. The sample temperature is measured using a
LS370 resistance bridge (Lake Shore). The heat sink is weakly
coupled to the mixing chamber and its temperature is PID controlled
with a heater within the field-compensated region of the large
superconducting magnet. In a typical measurement of the temperature
dependence of the MCE, we slowly sweep the temperature of the heat
sink leading to a slow sweep of the average sample temperature. Due
to the alternating magnetic field, the sample temperature oscillates
through its average value. The amplitude of this oscillation gives
the MEC $\Gamma_H$, while the average value determines $T$. 

 \begin{figure}
 \includegraphics{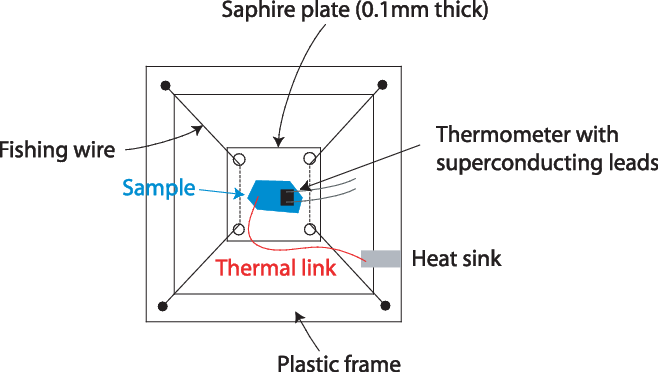}
 \caption{\label{}Schematic view of the platform used for
 alternating field magnetocaloric effect measurements under
 quasi-adiabatic conditions.}
 \end{figure}

\section{Application to strongly correlated materials}

Figure~3 shows a typical example of a temperature sweep during an
alternating field MCE measurement. We have chosen YbRh$_2$Si$_2$ as
this is a system close to a QCP, at which a pronounced divergence of
the magnetic Gr\"uneisen ratio has been
observed.\cite{ISI:000263389500045} The data discussed here are
obtained on a slightly Fe-doped single crystal of 20.1\,mg
mass in a shape of plate with a thickness of 0.25mm (residual resistivity ratio $\sim$12\cite{future_pub}). A relatively large piece of single crystal was chosen to determine precisely its magnetization and specific heat, which we will discuss later. The field was applied within the easy magnetic plane of the
system (parallel to the plane of the sample plate), which is perpendicular to the tetragonal $c$-axis. The
static field is set to 50\,mT which is the critical field for
undoped YbRh$_2$Si$_2$ and the alternating field has an amplitude of
4.5\,mT with frequency of 0.1\,Hz. A clear oscillation of the sample
temperature due to the MCE is resolved. The inset of Fig.~3(a)
displays the temperature variation on a larger time scale such that
individual oscillations could not be resolved. Here the thickness of
the line is a measure of the oscillation amplitude. Indeed the
thickness of the $T(t)$ trace increases as temperature is decreased,
indicating a larger temperature oscillation $dT/dH$ with decreasing
temperature. This is a clear evidence of quantum criticality in this
material, since $dT/dH$ is expected to decrease linearly with
decreasing temperature for a non-critical material with constant
magnetic Gr\"uneisen ratio $\Gamma_H=const$.

 \begin{figure}
 \includegraphics{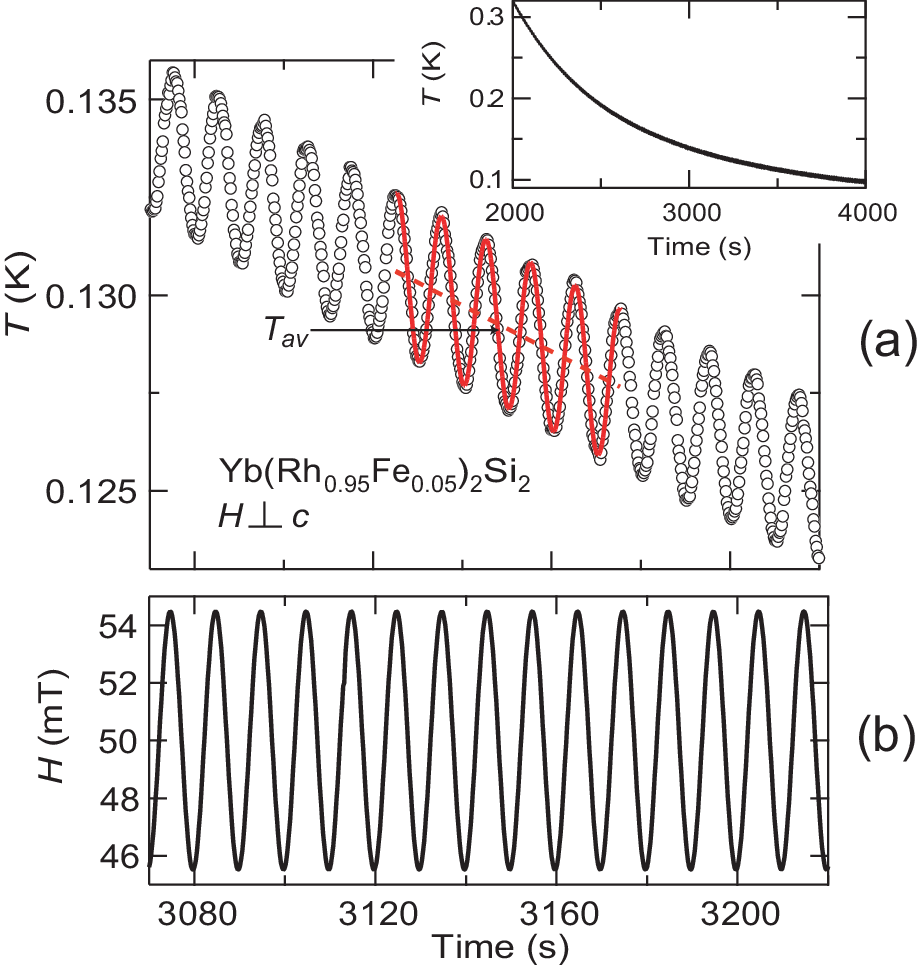}
 \caption{\label{}(Color online) Sample temperature (a) and magnetic field (b) vs time for a measurement
 on doped YbRh$_2$Si$_2$. The red solid and dotted lines show a fit to $T(t)=a_0+a_1 t+a_2 t^2+T_0\sin(2\pi f(t-\delta t))$ and its background term, $a_0+a_1 t+a_2 t^2$, respectively. The inset displays the temperature variation of the sample over
  a longer time scale.}
 \end{figure}

In order to quantify the MCE at a given average temperature
$T_{av}$, we chose time intervals of five periods (5$\times 1/f$
seconds in time) and determine the amplitude of the temperature
oscillation by fitting the temperature trace to $T(t)=a_0+a_1 t+a_2
t^2+T_0\sin(2\pi f(t-\delta t))$ (red solid line in Fig.~3(a)),
where $f$ is the frequency of the field modulation
$H(t)=H_{av}+H_0\sin(2\pi ft)$. The second order polynomial in time
($a_0+a_1 t+a_2 t^2$, cf. red dotted line in Fig.~3(a)) is added to
the fitting function to take into account the background temperature
drift due to the slow sweep of the heat sink temperature. The
amplitude of the temperature oscillation $T_0$ divided by the field
modulation amplitude $H_0$ gives the MCE. This value is associated
to the average temperature $T_{av}$ given by the background
temperature at the center of the time interval, $T_{av}=a_0+a_1
(t_0+2.5/f)+a_2 (t_0+2.5/f)^2$, where $t_0$ is the time of the first
data point for fitting. The magnetic Gr\"uneisen ratio is then
calculated by $\Gamma_H(T_{av})=1/T_{av}\times T_0/H_0$. By shifting
the fitting interval across the temperature trace a dense and
continuous curve of $\Gamma_H$ is obtained, as shown in Fig. 4. In
this measurement from 3~K down to 0.3~K, the temperature of the heat
sink was slowly decreased, using PID control, and below 0.3\,K the
heater for the heat sink was turned off, since the thermal
conductance of the thermal link is very small at low temperatures,
and thus, the decrease of the average sample temperature is very
slow. We have also performed a measurement with half value of the
frequency ($f$=0.05\,Hz) and the same $H_0$ in the entire
temperature range from 3 to 60\,mK, using the same cooling rate of
the heat sink, and confirmed that the result is exactly the same.
This proves that the measurement is taken under effective adiabatic
conditions ($1/f\ll\tau$; i.e., heat flow between sample and heat
sink within one period is negligible.). The data are compared with
$\Gamma_H(T)$ calculated using Eq. 1 from magnetization and specific
heat data~\cite{future_pub} taken at the same field of 0.05~T and
measured using the same piece of single crystal. The excellent
agreement between the two independent ways to obtain the magnetic
Gr\"uneisen ratio prove that this technique is best-suited to
quantitatively obtain the true adiabatic MCE. Furthermore, the
comparison shows the extraordinarily high resolution of the
alternating field technique.

 \begin{figure}
 \includegraphics{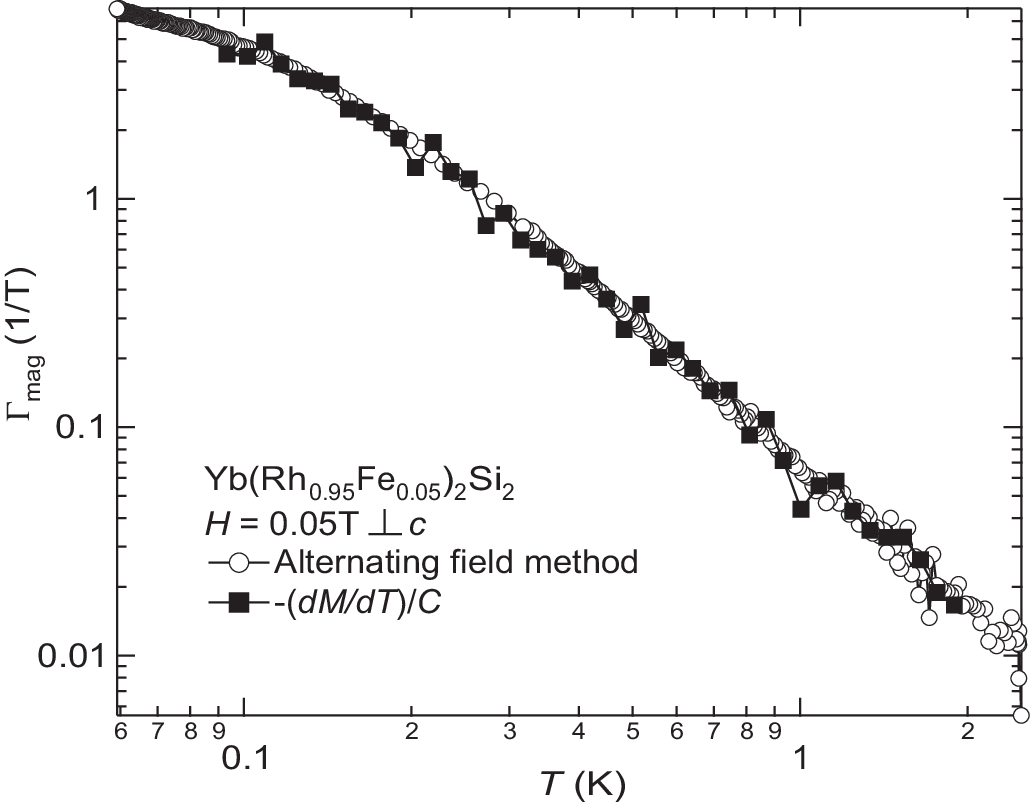}
 \caption{\label{}Magnetic Gr\"{u}neisen ratio for Yb(Rh$_{0.95}$Fe$_{0.05}$)$_2$Si$_2$ as
  a function of temperature. The open circles are determined by the alternating field
   technique while the solid squares are
   calculated from magnetization and specific heat data using $-(dM/dT)/C$.}
 \end{figure}

At last we discuss the problem of Joule heating caused by the field
modulation. If the sample is very clean and the mean free path of
the electrons is rather high, the frequency and the amplitude of
alternating field have to be low to avoid Joule heating due to eddy
currents. Figure~5 shows data obtained in the normal state of a
high-quality single crystal of the clean heavy-fermion
superconductor CeCoIn$_5$ in magnetic field applied parallel to
c-axis. Note here that the material has a very large electronic mean
free path of 810\AA\cite{movshovich:prl-01} and alternating-field is
applied perpendicular to the plane of a platelet sample
(1.3$\times$1.1$\times$0.33\,mm$^3$) with a twice larger frequency
of 0.2\,Hz than the one used for Fe-doped YbRh$_2$Si$_2$. This
condition causes a severe problem with eddy current heating. As
clearly seen in Fig.~5, the temperature of the sample oscillates
with twice the frequency of the field modulation which is due to
Joule heating. Eddy currents induced by the variation of a magnetic
field are proportional to the time derivative of the field,
$dH/dt=2\pi f H_0\cos(2\pi ft)$. Since eddy current heating is
proportional to the square of the current, $(dH/dt)^2\sim
f^2H_0^2[1+\cos(2\pi 2ft)]$, it results in an oscillation with $2f$
frequency. Note that the heating effect has two components; constant
and 2$f$-oscillating ones. Figure~5 also shows that as soon as the
alternating field is turned on, the average sample temperature
raises immediately due to the constant heating component, followed
by the 2$f$-oscillation. We have also verified that the amplitude of
$2f$ component increases rapidly with amplitude $H_0$ and frequency
of the alternating field, whereas the $1f$ component due to the
intrinsic MCE remains constant. The eddy current heating can very
effectively be suppressed by a reduction of the cross-section of the
sample perpendicular to the field axis and as well as by reducing
the frequency and/or amplitude of the field modulation.

 \begin{figure}
 \includegraphics{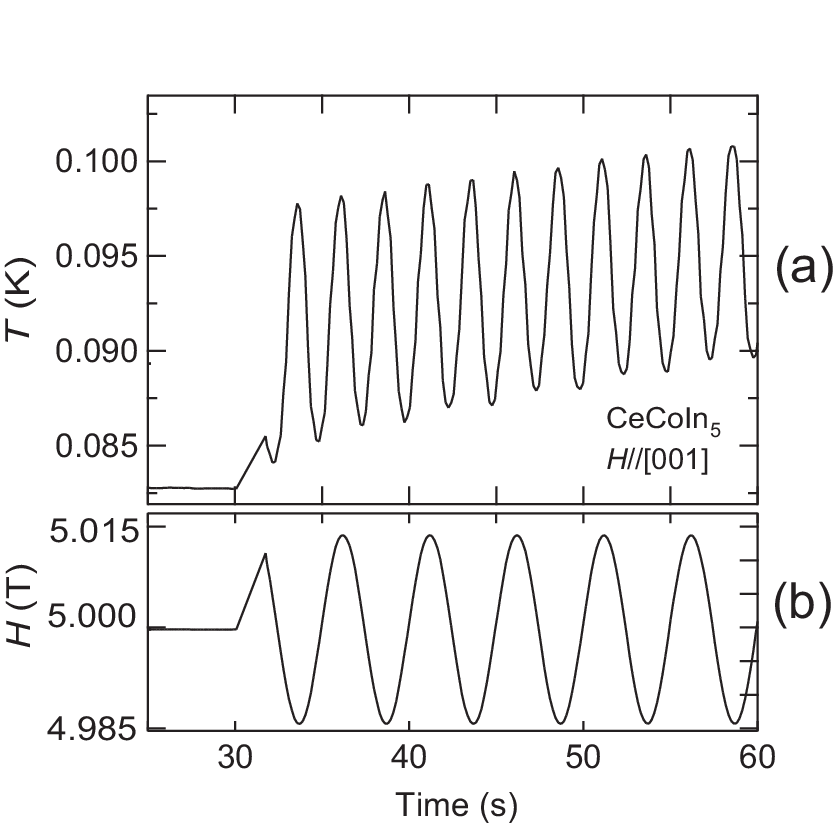}
 \caption{\label{} Sample temperature (a) and magnetic field (b) vs time for a high-quality
 single crystal of CeCoIn$_5$ for $H\parallel$[001]. Note that the temperature oscillation has twice
 the frequency of the field modulation.}
 \end{figure}

At an elevated temperature, where the electrical conductivity is
smaller, and therefore, the Eddy current heating is less severe, a
mixture of intrinsic 1$f$ and extrinsic 2$f$ oscillations can be
found, as shown in Fig.~6. The temperature as a function of time is
well fitted by a function with two oscillating components,
$T(t)=a_0+a_1 t+a_2 t^2+T_0\sin(2\pi f(t-\delta
t))+T_0^{2f}\cos(2\pi 2f(t-\delta t_{2f}))$ (red line in Fig.~6). We
deduced $T_0$, using several different frequencies but the same
$H_0$ and found a constant $T_0$. This indicates that the obtained
$T_0$ is still valid under the influence of eddy current heating.

 \begin{figure}
 \includegraphics{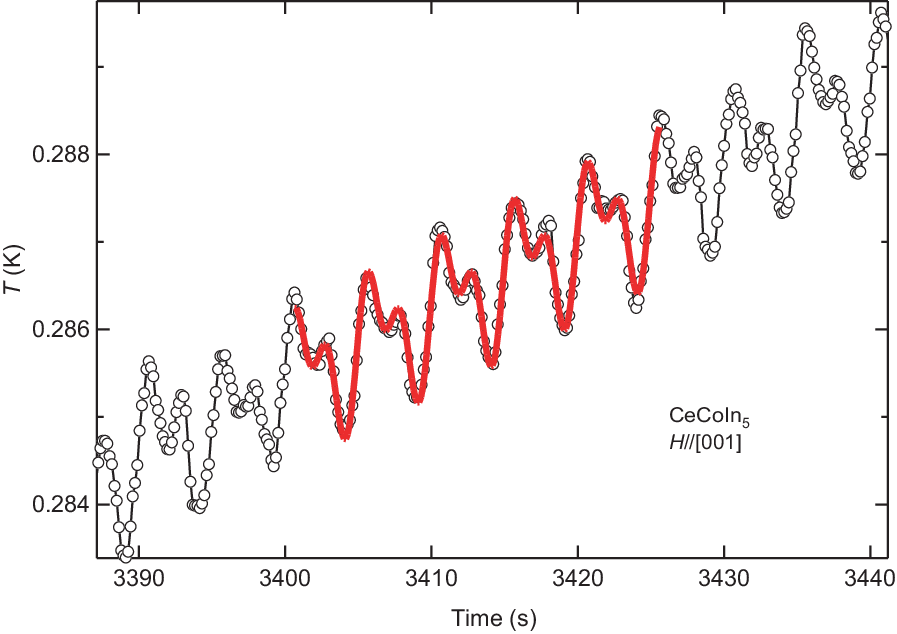}
 \caption{\label{}(Color online) Mixture of 1$f$- and 2$f$-oscillations of sample temperature of CeCoIn$_5$ under the same condition as Fig.~5. The red line is a fit to $T(t)=a_0+a_1 t+a_2 t^2+T_0\sin(2\pi f(t-\delta t))+T_0^{2f}\cos(2\pi 2f(t-\delta t_{2f}))$}
 \end{figure}



\section{Summary}
We developed a high-resolution low-temperature alternating-field
technique for quasi-adiabatic measurements of the MCE and magnetic
Gr\"uneisen ratio for clean metallic samples down to mK temperatures
and in large magnetic fields. The performance check has revealed
excellent agreement with the magnetic Gr\"uneisen ratio calculated
by magnetization and specific heat measurements. The main advantages
of the new technique compared to the latter are that it is much
faster and easier and that its resolution is much higher. The
capability of continuous temperature sweeps makes possible thorough
investigations of the magnetic Gr\"uneisen ratio in the entire
$H$-$T$ phase space, which is of particular interest for systems
close to magnetic-field induced QCPs.

\section{Acknowledgements}
We are grateful to M. Brando, M. Lang, R. Movshovic, A. Steppke, K.
Winzer and B. Wolf for valuable conversations. We thank H. S. Jeevan
and E. D. Bauer for providing the
Yb(Rh$_{0.95}$Fe$_{0.05}$)$_2$Si$_2$ and CeCoIn$_5$ single crystals,
respectively. Work supported by the German Science Foundation
through SFB 602 TPA19 and FOR 960 (Quantum phase transitions).


%
%

%




\begin{thebibliography}{8}

\bibitem{ZhuLijun:UnidGp}
L. Zhu, M. Garst, A. Rosch and Q. Si, Phys. Rev. Lett. \textbf{91}, 066404 (2003).

\bibitem{GarstM:SigctG}
M. Garst and A. Rosch, Phys. Rev. {\rm B} Cond. Matt. \textbf{72}, 205129 (2005).

\bibitem{KuchlerR:DivtGr}
R. Kuchler, N. Oeschler, P. Gegenwart, T. Cichorek, K. Neumaier, O. Tegus, C. Geibel, J.~A. Mydosh, F. Steglich, L. Zhu and Q. Si, Phys. Rev. Lett. \textbf{91}, 066405 (2003).

\bibitem{KuchlerR:Grurdq}
R. Kuchler, P. Gegenwart, K. Heuser, E.~W. Scheidt, G.~R. Stewart and F. Steglich,
 Phys. Rev. Lett. \textbf{93}, 096402 (2004).

\bibitem{Lorenz07}
T. Lorenz, S. Stark, O. Heyer, N. Hollmann, A. Vasiliev, A. Oosawa and H. Tanaka, J. Magn. Magn. Mater. \textbf{316}, 291 (2007).

\bibitem{ISI:000263389500045}
Y. Tokiwa, T. Radu, C. Geibel, F. Steglich and P. Gegenwart, Phys. Rev. Lett. \textbf{10} 066401 (2009).

\bibitem{ARostScience09}
A.~W. Rost, R.~S. Perry, J.-F. Mercure, A.~P. Mackenzie and S.~A. Grigera, Science \textbf{325}, 1360 (2009).

\bibitem{Fischer}
B. Fischer, J. Hoffmann, H.G. Kahle and W. Paul, J. Magn. Magn. Mater. \textbf{94}, 79 (1991).

\bibitem{future_pub}
Y. Tokiwa, M. Schubert, P. Gegenwart, J. S. Hirale, R. Movshovich, in preparation.

\bibitem{movshovich:prl-01}
R. Movshovich, M. Jaime, J.~D. Thompson, C. Petrovic, Z. Fisk, P.~G. Pagliuso and J.~L. Sarrao, Rev. Lett. \textbf{86}, 5152 (2001).

\end{thebibliography}
\end{document}